\documentclass[letterpaper,journal]{IEEEtran}
\usepackage{amsmath,amsfonts}
\usepackage{algorithmic}
\usepackage{array}
\usepackage[caption=false,font=normalsize,labelfont=sf,textfont=sf]{subfig}
\usepackage{textcomp}
\usepackage{url}
\usepackage{verbatim}
\usepackage{graphicx}
\usepackage{cite}
\usepackage{bm}

\def\BibTeX{{\rm B\kern-.05em{\sc i\kern-.025em b}\kern-.08em
    T\kern-.1667em\lower.7ex\hbox{E}\kern-.125emX}}
\usepackage{balance}

\begin{document}
\title{A CMOS+X Spiking Neuron with On-Chip Machine Learning}
\author{
    Steven Louis, Matthew Blake Abramson, Hannah Bradley, Cody Trevillian, Gene David Nelson,  Andrei Slavin,\\
    Artem Litvinenko, Jason Gorski, Ilya N. Krivorotov, Darrin Hanna, and Vasyl Tyberkevych
    \thanks{S. Louis  (SLouis@oakland.edu) and D. Hanna are with the Department of Electrical and Computer Engineering, Oakland University, Rochester, MI 48309, USA.}
    \thanks{H. Bradley, C. Trevillian, A. Slavin, A. Litvinenko, and V. Tyberkevych are with the Department of Physics, Oakland University, Rochester, MI 48309, USA.}
    \thanks{J. Gorski is with MicroNova, Waterford, MI 48328, USA.}
    \thanks{M. B. Abramson, I. N. Krivorotov, and G. D. Nelson are with the Department of Physics and Astronomy, University of California, Irvine, CA 92697, USA.}
}


\maketitle	
\begin{abstract}
We present the design and numerical simulation of a spiking neuron in a proof-of-concept model of on-chip machine learning. 
Built within the CMOS+X framework, the spiking neuron consists of an NMOS transistor combined with a magnetic tunnel junction (MTJ).
This NMOS+MTJ unit, when simulated in the industry-standard circuit simulation software ``LTspice'', reproduces multiple functions of a  biological neuron, including threshold spiking, latency, refractory periods, synaptic integration, and inhibition. 
These behaviors arise from the intrinsic magnetization dynamics of the MTJ and do not require any additional control circuitry.
By interconnecting the NMOS+MTJ neurons, we construct a model of an analog multilayer network that learns through spike-timing-dependent weight updates derived from a gradient-descent rule, with both training and inference modeled in the analog domain.
The simulated network demonstrates spike propagation and successful training on the XOR task under the modeled conditions.
\end{abstract}

\begin{IEEEkeywords}
Neuromorphic computing, Magnetic tunnel junctions (MTJ), Beyond CMOS, CMOS+X, CMOS-X, NMOS+MTJ, Spiking neural networks (SNN), Analog circuits, Gradient descent, On-chip learning, On-chip training, Spintronics, Hardware neural networks, Edge computing, Analog computation, CMOS integration, In-memory computing, Synaptic integration, leaky integrate and fire (LIF).
\end{IEEEkeywords}

\section{Introduction}

\IEEEPARstart{M}{odern} edge-computing platforms face growing demands for real-time intelligence under strict constraints on energy, latency, and silicon area \cite{li2019edge}. 
Conventional deep-learning architectures rely on dense matrix operations and high-precision arithmetic, which make them computationally expensive and poorly suited for battery-powered or thermally limited devices \cite{jouppi2017datacenter}. 
In contrast, spiking neural networks (SNNs) offer an alternative that can reduce energy consumption through sparse, event-driven computation, in which information is carried by discrete spikes and computation occurs only when spikes are present \cite{maass1997networks, davies2018loihi}. 
These advantages have made SNNs increasingly attractive for embedded and edge-AI applications, where the energy-efficiency is as important as the accuracy.

Achieving these benefits in hardware requires moving beyond traditional digital architectures \cite{markovic2020physics}. 
However, the implementation of \textit{digital} SNNs faces memory bottlenecks, clock-synchronous overhead, and substantial energy per operation \cite{tang2023seneca}. 
As a result, there is growing interest in analog, in-memory neuromorphic computing, where the nanoscale  analog devices perform neural computation directly \cite{markovic2020physics, tang2023seneca, grollier2016bioinspired, finocchio2024roadmap, Incorvia, indiveri2015memory}. 
Such systems can exploit device physics to realize parallelism and small circuit footprints, positioning neuromorphic hardware as a promising approach for scalable and potentially energy-efficient AI systems \cite{indiveri2015memory}. 

CMOS+X is a recently proposed hardware framework designed to meet these needs by integrating standard CMOS with emerging analog devices for neuromorphic computing \cite{Incorvia, nsf2022cmosx}.
This device-agnostic architecture incorporates beyond-CMOS technologies, including spintronic, memristive, and photonic components, which can be used to realize compact spiking neurons and tunable synapses within an analog, CMOS-compatible environment. 
For many CMOS+X approaches, it is desirable for device-level neural elements to generate spiking behavior and to participate in learning computations without relying on digital control or on a  separation between memory and computation.

In this work, we investigate a spintronic magnetic tunnel junction (MTJ) as the manufacturable ``X'' element for the CMOS-X neuromorphic processors, demonstrated here in simulation to support both simulated training and inference. 
MTJs offer room temperature operation, tunable resistance, well-understood spin-transfer-torque (STT) dynamics, and established fabrication processes already deployed in embedded MRAM technology, making MTJs highly suitable for integration with CMOS \cite{chappert, alzate, edelstein, everspin, sun}. 
Prior work has shown that MTJs can emulate neuronal behavior through their magnetization dynamics \cite{mtjneuron}, and numerous studies have demonstrated their potential in neuromorphic and unconventional computing systems \cite{markovic2020physics, grollier2016bioinspired, finocchio2024roadmap, Incorvia, salinas2023lastmile, chumak2022roadmap, jiang2024spin}.

We demonstrate below that pairing an MTJ with a single NMOS transistor, forming a structure nearly identical to a standard 1T-1MTJ MRAM cell \cite{Lin2009}, naturally yields spiking behavior through STT-driven magnetization dynamics. 
The resulting NMOS+MTJ neuron reproduces essential neuronal properties, including threshold activation, spike generation, latency, refractory behavior, synaptic integration, and inhibition.
These properties emerge naturally from the coupled device physics and transistor operation, enabling development of artificial neurons compatible with analog CMOS circuitry and standard semiconductor manufacturing. 

We further demonstrate that simulated networks of NMOS+MTJ neurons, interconnected by analog synapse models, can perform feedforward inference and gradient-descent learning within LTspice using time-encoded spike signals.
In our model system, each neuron operates as a single-fire spiking unit, a structure that enables differentiable spike timing and makes gradient-descent-based learning feasible within an analog SNN framework. 
Using simulations in LTspice, we construct a multilayer spiking neural network capable of learning the nonlinear XOR classification task, used here as a simple nonlinear proof-of-concept demonstration, entirely within the analog domain. 
This includes an analog implementation of supervised learning using spike-timing-based gradient descent.

This paper proceeds as follows. 
Section II reviews the electrical and magnetic properties of MTJs relevant to neuromorphic operation. 
Section III demonstrates how a single NMOS+MTJ pair functions as a spiking neuron and reproduces multiple biologically inspired behaviors. 
Section IV describes how these neurons can be interconnected through analog synapses to form functional multilayer networks. 
Section V presents an analog, gradient-descent learning demonstration on the XOR task. 
Together, these results establish a proof-of-concept for trainable, CMOS-compatible neuromorphic circuits based on MTJ-enabled spiking neural networks.

\section{Magnetic Tunnel Junctions}

\subsection{Neuron configuration}
Magnetic tunnel junctions and spin valves are nanoscale spintronic devices that have been used in data storage technologies for two decades \cite{chappert}.  
These devices were first used as read heads in hard disk drives, and later were adopted as nonvolatile memory elements in magnetoresistive random access memory (MRAM) \cite{alzate, edelstein, everspin}.  
A typical MTJ design consists of a multilayer structure composed of alternating ferromagnetic and nonmagnetic layers.  
The operation relies on the tunneling magnetoresistance effect, in which the electrical resistance of the MTJ depends on the relative orientation of the magnetization directions in the ferromagnetic layers \cite{sun}.  

A recent theoretical study has shown that under specific operating conditions, an MTJ can exhibit electrical behaviors that are similar to those of biological neurons \cite{mtjneuron}.  
These behaviors include threshold based spiking, integration of input signals over time, refractory periods, and inhibition, which are all features observed in biological neurons.  
These behaviors arise from the magnetization dynamics of the MTJ and do not require any additional control circuitry. 

\begin{figure}
\centering
\includegraphics{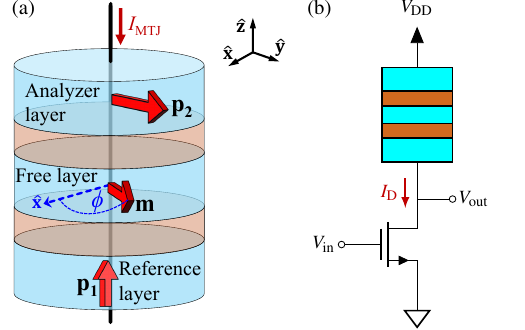}
\caption{
(a) MTJ structure showing the reference layer $\mathbf{p}_1$, the analyzer layer $\mathbf{p}_2$, and the free layer $\mathbf{m}$ whose magnetization can rotate in the $xy$-plane.  
The resistance $R_{\mathrm{MTJ}}(\phi)$ depends on the angle between $\mathbf{m}$ and $\mathbf{p}_2$.  
(b) NMOS+MTJ neuron circuit in which a magnetic tunnel junction is connected in series with an NMOS transistor.  
The input voltage $V_{\mathrm{in}}$ is applied to the transistor gate, and the output voltage $V_{\mathrm{out}}$ is measured at the drain.  
The series connection ensures that the same current $I_{\mathrm{MTJ}}$ equals the drain current $I_D$ and flows through both elements.  
}
\label{toon}
\end{figure}

The MTJ used in this work consists of multiple stacked magnetic and dielectric layers, as shown in Figure \ref{toon}(a).  
Two of these layers have fixed magnetization directions, set by their internal anisotropy and pinning structure. These layers are referred to as the reference layer ($\mathbf{p}_1$) and the analyzer layer ($\mathbf{p}_2$).
Between these layers is the free layer, whose magnetization $\mathbf{m}$ is not fixed and can rotate in response to external influences \cite{mtjneuron}.
MTJ stacks implemented in commercial technologies are more complex than this model and often include features such as synthetic antiferromagnet-pinned fixed layers and low-anisotropy free layers; however, the simplified model used here is sufficient for this work.

The resistance of the MTJ, $R_\mathrm{MTJ}(\phi)$, depends on the angle between $\mathbf{m}$ and $\mathbf{p}_2$ \cite{sun}, where $\phi$ is defined as the angle between $\mathbf{m}$ and the x-axis.  
In this work, the analyzer layer $\mathbf{p}_2$ is assumed to be fixed in the $\mathbf{\hat{y}}$ direction, as shown in the coordinate system in Figure \ref{toon}.  
The magnetization vector $\mathbf{m}$ is assumed to remain approximately within the $\mathbf{\hat{x}\hat{y}}$-plane \cite{mtjneuron}.  
A small external magnetic field is applied in the $\mathbf{\hat{x}}$ direction so that in the absence of an external disturbance, $\mathbf{m}$ is oriented in the $\mathbf{\hat{x}}$ direction.  
  
The relationship between $\phi$ and the MTJ resistance is as follows.  
When $\phi = 90^\circ$, $\mathbf{m}$ and $\mathbf{p}_2$ are aligned in parallel, $R_\mathrm{MTJ}$ is at a minimum.  
In this work, the resistance in the parallel state is assumed to have a value of $R_P = 500~\Omega$.  
When $\phi = 270^\circ$, $\mathbf{m}$ and $\mathbf{p}_2$ are in antiparallel alignment, and $R_\mathrm{MTJ}$ is at a maximum.  
The resistance in the antiparallel state is assumed to have a value of $R_{AP} = 1500~\Omega$.  
These values for $R_P$ and $R_{AP}$ are typical for MTJs.  
For other values of $\phi$, the $R_\mathrm{MTJ}$ varies between $R_P$ and $R_{AP}$ (eq. (1) in \cite{phyBased}).  

The magnetization of the reference layer $\mathbf{p}_1$ is oriented in the $\mathbf{\hat{z}}$ direction.  
When a current $I_\mathrm{MTJ}$ flows through the MTJ, some of the electrons in the current become spin polarized by aligning their spins with the magnetization of $\mathbf{p}_1$.  
As these spin polarized electrons reach the free layer, they apply a torque to $\mathbf{m}$ through a physical effect known as spin transfer torque (STT) \cite{slon1, slon2}.  
This torque causes $\mathbf{m}$ to evolve over time \cite{krivorotov2008time}.  
Therefore, the dynamics of $\mathbf{m}$, and thus the angle $\phi$ and indirectly $R_\mathrm{MTJ}(\phi)$, are influenced by $I_\mathrm{MTJ}$.  
Please note that the STT from the analyzer layer is negligible because $\mathbf{p}_2$ is oriented in-plane, and therefore, it does not significantly affect the free-layer dynamics \cite{mtjneuron}.
In addition, the resistance of the $\mathbf{p}_1$--free-layer junction is negligible compared with the TMR resistance of the $\mathbf{p}_2$--free-layer junction, because the $\mathbf{p}_1$--free-layer spacer is metallic whereas the $\mathbf{p}_2$--free-layer spacer is an insulating tunnel barrier \cite{mtjneuron}.

The STT effect is responsible for switching memory states in MRAM \cite{worledge}.  
In MRAM, a current pulse of sufficient magnitude and duration changes the relative orientation of the free and fixed magnetic layers, switching the MTJ resistance between the parallel and antiparallel states~\cite{Bhatti2017,Nguyen2024}.  
This process is repeatable, non-volatile, and operates within standard CMOS circuit environments.
These characteristics have enabled MRAM to become a commercially available embedded memory technology \cite{everspin}.  
Consequently, the MTJ-based circuits described in the following sections can, in principle, be realized within standard semiconductor manufacturing flows that already support MRAM.

\subsection{MTJs and AI}

Our work seeks to use MTJs in analog circuits to perform machine learning.
This work builds upon a rapidly expanding body of research that seeks to implement machine learning functions directly in spintronic hardware. 
This section provides a concise overview of representative studies at the intersection of spintronics, artificial intelligence, and machine learning.  
This brief survey is intended to orient readers who are new to spintronics and highlight recent experimental and theoretical studies.  
For broader coverage, see these reviews on neuromorphic and unconventional spintronic computing \cite{markovic2020physics, grollier2016bioinspired, finocchio2024roadmap, Incorvia, salinas2023lastmile, chumak2022roadmap, jiang2024spin}. 
Because the cited studies differ substantially in device type, architecture, application, and level of physical implementation, the following is qualitative and is intended to position the present work within the broader field rather than to establish a quantitative performance comparison.

Experimental studies have demonstrated that magnetic nanodevices can directly implement machine-learning functions.  
For instance, a single MTJ has been used for temporal pattern classification in reservoir computing applied to spoken-digit recognition~\cite{torrejon2017}, and a similar experiment showed single-MTJ reservoir computing for temporal signal classification~\cite{riou2019}.  
Arrays of coupled MTJs have also been used for pattern recognition, such as vowel classification using synchronized dynamics in a four-MTJ network~\cite{romera2018}.  
Another study achieved image classification with a $4 \times 2$ MTJ crossbar array~\cite{Zhou2021}. 
MTJs have also been used to implement stochastic neuron behavior, including MTJ-based neural networks \cite{kaiser2022hardware}.
Larger systems, such as a $15 \times 15$ STT-MTJ crossbar array programmed with binary weights, have experimentally demonstrated inference on the Wine dataset, validating the feasibility of MTJ-based hardware classifiers~\cite{Goodwill2022}.  
Together, these results establish the experimental foundation for spintronic implementations of machine-learning architectures.  

Experimental work in MRAM has also demonstrated its potential for both analog and digital computation in neural architectures.  
Multi-level resistance states in MRAM cells confirm stable switching for analog synaptic weighting~\cite{Rzeszut2022}.  
An embedded memory array containing over eight million MTJs was used not only to store quantized synaptic weights as nonvolatile resistance states, but also to perform multiply--accumulate (MAC) operations directly within the MTJ array~\cite{you20238b}.  
MRAM-based arrays have performed logic and arithmetic operations, directly within memory~\cite{Lv2024}.  
Analog multiply–accumulate behavior in MRAM crossbars was examined experimentally in~\cite{Kiseleva2025}, while mixed-signal MRAM–CMOS circuits demonstrated real-time neural emulation with multi-level analog synapses and voltage-domain processing~\cite{Lee2025}.  

The neuron used in this work was described previously in~\cite{mtjneuron}, where it was shown that MTJ dynamics can follow the artificial neuron equation (ANE).  
Similar ANE equations have been derived for a wide range of devices, including antiferromagnetic neurons~\cite{Khymyn}, synthetic antiferromagnetic oscillators~\cite{Liu}, optically initialized antiferromagnetic heavy-metal heterostructures~\cite{Mitro}, spin Hall nano-oscillators~\cite{Ovcha}, and Josephson junctions~\cite{Crotty2010,Schneid}. 
Thermal-fluctuation–assisted MTJ neurons were also simulated in~\cite{rodrigues2023spintronic}, and additional MTJ-based neuron implementations have been reported in~\cite{zhang2025dual}.
A review of neurons governed by the ANE was presented in~\cite{Bradley2023}, which summarized the theoretical characteristics common to these devices.  
Since that publication, two studies have demonstrated machine-learning applications based on the ANE~\cite{Bradley2024,Sotnyk2025}, and yet another has shown that the same formalism can reproduce a biological withdrawal reflex and can be mapped onto the Hodgkin–Huxley model~\cite{BradleyWithD}.  

While these studies established that spintronic devices can support neuromorphic computation, much of the relevant information presented in the existing literature remains fragmented between device-level demonstrations \cite{torrejon2017, riou2019, romera2018, Zhou2021, kaiser2022hardware, Goodwill2022}, memory-centric neural-network architectures \cite{Rzeszut2022,you20238b, Lv2024, Kiseleva2025, Lee2025}, and biologically inspired neuron models \cite{mtjneuron, Khymyn,Liu, Mitro,  Ovcha, Crotty2010,Schneid, rodrigues2023spintronic, zhang2025dual, Bradley2023, Bradley2024,Sotnyk2025, BradleyWithD}. 
The studies reviewed above employ MTJs in several roles, including as nonvolatile memory elements and neuron-like devices within broader circuit architectures. 
The present work focuses specifically on how the intrinsic temporal dynamics of the MTJ can participate in computation and learning within a larger analog CMOS circuit framework.

The objective of this work is to investigate whether neuromorphic CMOS+MTJ analog circuitry can directly support supervised gradient-based learning using device structures compatible with modern semiconductor fabrication. 
In the proposed framework, biologically inspired neuron behaviors emerge from the coupled dynamics of the MTJ and transistor, rather than from the externally imposed mathematical neuron models. 
To the best of our knowledge, this is the first circuit-level neuromorphic hardware simulation in which such physically generated neuron dynamics participates directly in supervised learning within a spiking neural network.

\section{NMOS+MTJ Spiking Neuron \label{sectmtjnmos}}

The present work combines an MTJ with a single NMOS transistor, as shown in Figure~\ref{toon}(b), to form a functional artificial neuron.  
This configuration uses the MTJ to generate spikes through STT driven magnetization dynamics, and uses the NMOS transistor to control the current that drives these dynamics.  
Both the input to the neuron and the output from the neuron are voltages, which makes the circuit compatible with conventional CMOS voltage signals.  
The combination produces a compact neuromorphic circuit element that can be used as a building block for construction of neuromorphic networks.  
In this section, the circuit configuration, operating principle, and simulation results for the NMOS+MTJ neuron are presented.  

For the electrical neuron shown in Figure~\ref{toon}(b), the input voltage $V_{\mathrm{in}}$ is applied to the gate of the NMOS transistor, and the output voltage $V_{\mathrm{out}}$ is measured at the drain.  
The output is given by the expression  
\begin{equation}  
V_{\mathrm{out}} = V_{\mathrm{DD}} - I_D R_{\mathrm{MTJ}}(\phi)  \label{ohms}  
\end{equation}  
Here, $V_{\mathrm{DD}}$ is the supply voltage, $R_{\mathrm{MTJ}}(\phi)$ is the resistance of the MTJ, and $I_D$ is the drain current of the transistor.  
The MTJ and the NMOS transistor are connected in series, so that the same current flows through both elements, that is, $I_D = I_{\mathrm{MTJ}}$.  
The gate voltage $V_{\mathrm{in}}$ determines the value of $I_D$, which controls the neuron response.  
Together, the MTJ and NMOS transistor form what is referred to in this paper as an ``NMOS+MTJ neuron''.  

It is worth noting that the circuit in Figure~\ref{toon}(b) is nearly identical to a one-transistor-one-MTJ (1T-1MTJ) cell commonly used in MRAM \cite{Lin2009}.  
In a conventional 1T-1MTJ cell, the MTJ serves as a non-volatile storage element whose resistance encodes a binary state, while the access transistor controls read and write operations.  
By adapting this familiar structure for dynamic operation, the present design uses device technology compatible with existing manufacturing. 

The circuit shown in Figure~\ref{toon}(b) provides the foundation for this work.
In this section, we demonstrate that this cell can reproduce behaviors that are characteristic of biological neurons.
In the following sections, we demonstrate that the NMOS+MTJ neuron can be interconnected into spiking neural networks with tunable synapses capable of performing gradient-descent-based learning in the analog domain.

Simulations of this circuit were carried out using LTspice, a widely used tool for analog circuit simulation \cite{ltspice}.
The model of an NMOS transistor was selected from the default component library included with the software.
The simulations in this work used the default NMOS transistor model included in the LTspice program package without process-specific optimization of its parameters.
Consequently, the operating voltage was not optimized for scaled CMOS technologies.  
The MTJ model was adapted from a previously published design \cite{phyBased}, with details of the modifications and the simulation parameters provided in Appendix~A. 
The present simulations use an isothermal MTJ model. 
They do not include process variation, device mismatch, temperature-dependent resistance, thermal fluctuations, or self-heating effects.
Accordingly, the present work does not establish robustness to device or operating-condition variations.

\begin{figure}
\centering
\includegraphics{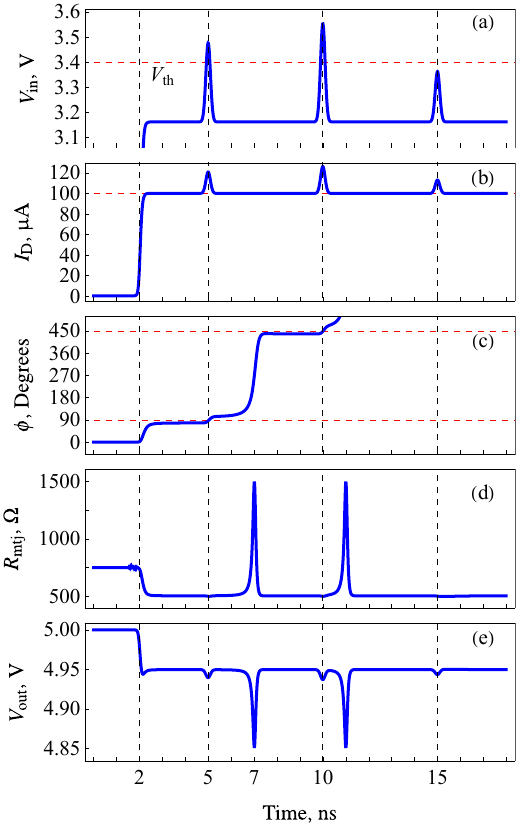}
\caption{
Simulation results of the NMOS+MTJ neuron circuit in LTspice.  
(a) Input gate voltage $V_{\mathrm{in}}$.  
(b) Corresponding drain current $I_D$.  
(c) Free layer magnetization angle $\phi$.  
(d) Time-dependent MTJ resistance $R_{\mathrm{MTJ}}$.  
(e) Output voltage $V_{\mathrm{out}}$.  
}
\label{demo}
\end{figure}

Figure~\ref{demo} shows the results of an LTspice simulation demonstrating NMOS+MTJ operation.  
At the beginning of the simulation, the gate voltage is set to $V_\mathrm{in} = 0~\mathrm{V}$, which places the NMOS transistor in cutoff and prevents current from flowing through the circuit (Figure~\ref{demo}(b)).  
Under these conditions, the magnetization vector $\mathbf{m}$ is aligned with the external magnetic field in the $\mathbf{\hat{x}}$ direction, resulting in an angle $\phi$ equal to $0^\circ$ (Figure~\ref{demo}(c)).  
At this angle, the MTJ maintains a resistance of $750~\Omega$ (Figure~\ref{demo}(d)), and the output voltage is equal to the supply voltage $V_{\mathrm{DD}} = 5~\mathrm{V}$ (Figure~\ref{demo}(e)).  

At time $t = 2~\mathrm{ns}$, the gate voltage increases to a bias voltage $V_{\mathrm{bias}} = 3.16~\mathrm{V}$ (Figure~\ref{demo}(a)), placing the NMOS transistor in saturation with a drain current $I_D$ of approximately $100~\mu\mathrm{A}$ (Figure~\ref{demo}(b)).  
This current applies an STT that rotates the magnetization vector $\mathbf{m}$ in the free layer from $\phi = 0^\circ$ to approximately $80^\circ$, bringing $\mathbf{m}$ close to alignment with $\mathbf{p}_2$ (Figure~\ref{demo}(c)).  
As a result, the MTJ resistance decreases to about $500~\Omega$, and the output voltage, which can be calculated from Eq.~(\ref{ohms}), drops to approximately $V_{\mathrm{out}} = 4.95~\mathrm{V}$.  

At time $t = 5~\mathrm{ns}$, the input voltage $V_{\mathrm{in}}$ briefly increases to $3.45~\mathrm{V}$.  
This short pulse causes the drain current $I_D$ to briefly increase to $120~\mu\mathrm{A}$ before returning to $100~\mu\mathrm{A}$.  
The brief increase in current, together with the instantaneous value of $R_{\mathrm{MTJ}}$, produces a small bump at $t=5$ ns in the output voltage $V_{\mathrm{out}}$, as shown in Figure~\ref{demo}(e).  
A more important change also occurs at this time.  
The elevated current increases the STT and drives the magnetization angle $\phi$ beyond $90^\circ$.  
After a delay of about $2~\mathrm{ns}$, the free layer magnetization $\mathbf{m}$ completes a full $360^\circ$ precession at $t = 7~\mathrm{ns}$ and then stabilizes at $\phi = 440^\circ$, which is equivalent to $80^\circ \bmod 360$.  
During this rotation, the MTJ resistance $R_{\mathrm{MTJ}}$ rapidly increases from $500~\Omega$ to $1500~\Omega$ and then returns to $500~\Omega$.  
This sharp and temporary change in resistance causes a distinct peak in the output voltage with a minimum value of $4.85~\mathrm{V}$, as shown in Figure~\ref{demo}(e).  
In the context of spiking neural networks, this type of output waveform is referred to as a ``spike'' because it resembles the voltage spike of biological neurons. 
The ability of the MTJ–NMOS pair to generate such spikes demonstrates its suitability for use as a spiking artificial neuron.

The rotation of the free layer magnetization can be explained by the existence of a threshold angle $\phi_{\mathrm{th}}$.  
With only the bias current applied, the magnetization remains near $\phi = 80^\circ$.  
If $\mathbf{m}$ momentarily rotates past $\phi_{\mathrm{th}}$, the spin transfer torque drives a complete precession of the magnetization.  
For the system used in this work, the threshold angle is $\phi_{\mathrm{th}} = 90^\circ$.  
The value of the input voltage $V_{\mathrm{in}}$ required to push $\phi$ past this angle represents the circuit-level threshold condition, and it depends on the combination of the bias voltage $V_{\mathrm{bias}}$ and the instantaneous impulse voltage applied at the gate.  
For the bias condition of $V_{\mathrm{bias}} = 3.16~\mathrm{V}$ used here, a single input spike that raises $V_{\mathrm{in}}$ above approximately $V_{\mathrm{th}} = 3.4~\mathrm{V}$ is sufficient to exceed $\phi_{\mathrm{th}}$.  
As shown by the dashed line in Figure~\ref{demo}(a), the input pulse applied at $t = 5~\mathrm{ns}$ exceeds $V_{\mathrm{th}}$ and therefore initiates the full $360^\circ$ precession observed in the simulation.  

Additional simulations were performed with varied MTJ resistance ranges under the same input waveform.  
The drain current and the free-layer trajectory $\phi(t)$ remained unchanged, as the intrinsic spiking dynamics is governed by the current-driven spin-transfer torque and are independent of the resistance.  
Consequently, the resistance variation affects the output voltage through Eq.~(\ref{ohms}), rather than the spiking behavior. 

At time $t = 10~\mathrm{ns}$, there is a second input impulse with an amplitude of $3.55~\mathrm{V}$, which is larger than the amplitude of the first impulse.  
In other words, the input voltage $V_{\mathrm{in}}$ is even further above the threshold voltage $V_{\mathrm{th}}$ than in the first case.  
The larger input momentarily produces a stronger STT that causes the magnetization to rotate past the threshold angle $\phi_{\mathrm{th}}$ more quickly.  
As a result, the delay between the input pulse and the onset of precession is reduced from about $2~\mathrm{ns}$ to about $1~\mathrm{ns}$.  
Once a full $360^\circ$ precession is initiated, the free layer magnetization follows the same trajectory as before, starting from its initial stationary position, passing through antiparallel alignment, and returning to its starting orientation.  
Because the amplitude of the MTJ resistance spike depends on the total angular excursion rather than the speed at which the precession begins, the output voltage $V_{\mathrm{out}}$ retains the same spike shape regardless of how quickly the precession is initiated.  

At time $t = 15~\mathrm{ns}$, there is a third input impulse with an amplitude of $3.35~\mathrm{V}$, which is below the threshold voltage $V_{\mathrm{th}}$.  
This input causes a small increase in the drain current, however, the STT is not sufficient to initiate a full rotation of the free layer magnetization.  
As a result, no spike is generated, and the output voltage $V_{\mathrm{out}}$ remains unchanged.  

Together, the responses of the NMOS+MTJ neuron to these three input perturbations demonstrate an important feature observed in biological neurons, known as threshold activation.  
Threshold activation, also referred to as the all-or-nothing response, means that the neuron remains inactive if the input is below the threshold condition.  
If the input exceeds the threshold, the neuron generates a spike with the same characteristic shape regardless of the strength of the input.  
In the NMOS+MTJ neuron, the threshold condition corresponds to the magnetization angle $\phi$ exceeding the threshold angle $\phi_{\mathrm{th}}$.  
If $\phi$ remains below $\phi_{\mathrm{th}}$, the neuron stays inactive and no spike is generated.  
If $\phi$ exceeds $\phi_{\mathrm{th}}$, the neuron generates a spike with the same characteristic shape regardless of how quickly $\mathbf{m}$ moves beyond the threshold angle.  

The simulation also demonstrates a second feature observed in biological neurons, known as response latency.  
Response latency is the measurable delay between the application of an input and the onset of spiking activity \cite{levakova2015review}.  
As in biological neural systems, the latency in the NMOS+MTJ neuron depends on the strength of the input.  
Stronger inputs result in shorter delays, while inputs close to the threshold condition produce longer delays before a spike occurs. 
This was observed in Figure~\ref{demo} at $t=5,~10$ ns.

\begin{figure}
\centering
\includegraphics{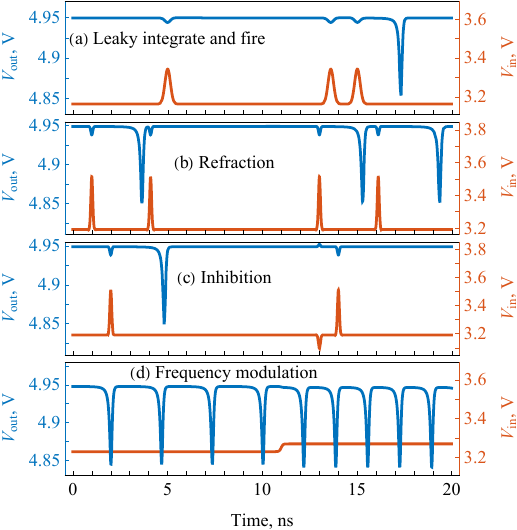}
\caption{Biological neuron-like behaviors reproduced by the NMOS+MTJ neuron circuit.  
In the plot, the orange curve shows the input voltage and the blue curve shows the output voltage.
(a) Leaky integrate and fire / synaptic integration: multiple subthreshold inputs combine over time to trigger a spike.
(b) Refractory period: demonstrates absolute and relative refraction following an initial spike.
(c) Inhibition: an inhibitory signal prevents spiking that would otherwise occur.
(d) Frequency modulation: constant input above threshold produces a spike train whose frequency depends on input amplitude.
} 
\label{bio}
\end{figure}

Figure~\ref{bio} presents additional simulations of neuron-like behaviors reproduced by the NMOS+MTJ neuron under a variety of excitation conditions.  
In all the considered cases, the input signals consist of Gaussian-shaped voltage pulses; but the pulse amplitudes, widths, time delays, and bias conditions are intentionally varied to demonstrate different dynamical responses of the artificial neuron.  
These simulations illustrate how the coupled MTJ and NMOS dynamics can reproduce behaviors commonly associated with biological neurons.
In larger networks, the corresponding inputs would be supplied by upstream NMOS+MTJ neurons rather than by externally applied Gaussian pulses.

We next consider a simulation that examines synaptic integration in the NMOS+MTJ neuron.  
While the previous results focused on threshold activation and response latency using single input pulses, this simulation demonstrates how the neuron responds to multiple subthreshold inputs applied in close succession.  
Synaptic integration is an important computational process in biological neurons in which multiple excitatory and inhibitory inputs are combined over time to determine whether the membrane potential reaches the firing threshold.  
This capability allows biological neurons to detect temporal patterns, perform coincidence detection, and encode complex spatiotemporal information.  

In neuromorphic computing, temporal aspects of synaptic integration are commonly approximated using the leaky integrate-and-fire (LIF) model, in which the membrane potential integrates incoming signals while gradually decaying toward a resting value, and a spike is produced once the threshold is crossed \cite{gerstner2002spiking}.  
Figure~\ref{bio}(a) illustrates that the NMOS+MTJ neuron exhibits this same functional behavior.  
In the figure, the orange curve represents the input voltage and the blue curve represents the output voltage.  
When a single subthreshold input spike is applied at $t = 5~\mathrm{ns}$, the neuron remains inactive and no spike is generated.  
When two subthreshold spikes arrive in succession at $t = 15~\mathrm{ns}$, their combined effect pushes the magnetization angle $\phi$ past the threshold angle $\phi_{\mathrm{th}} = 90^\circ$, producing an output spike.  
This result confirms that the NMOS+MTJ neuron can perform synaptic integration in a manner consistent with the LIF model.  

Figure~\ref{bio}(b) illustrates a refractory response, a fundamental behavior observed in biological neurons \cite{gerstner2002spiking}.  
The absolute refractory period is the short interval immediately following a spike during which no subsequent firing can occur.  
In the LTspice simulation, an input pulse at $t = 1~\mathrm{ns}$ generates a spike, while a pulse of equal amplitude at $t = 4.1~\mathrm{ns}$ (3.1 ns between inputs) fails to produce an output spike.  
This occurs because the free-layer magnetization has not yet recovered sufficiently from the previous precessional event to again cross the spiking threshold.  

The relative refractory period is the interval during which an additional spike can occur, but with altered response characteristics due to incomplete recovery of the neuron state.  
In the LTspice simulation, an input pulse at $t = 13~\mathrm{ns}$ generates a spike, and a subsequent pulse at $t = 16.2~\mathrm{ns}$ (3.2 ns between inputs) also produces a spike.  
In this case, the free-layer magnetization had recovered sufficiently to cross the spiking threshold, but had not fully relaxed to its equilibrium orientation, thereby modifying the time required for the magnetization trajectory to produce the output spike.

Figure~\ref{bio}(c) illustrates inhibition, another important feature of biological neurons.  
At $t = 2~\mathrm{ns}$, an uninhibited input pulse with an amplitude above the firing threshold of $V_{\mathrm{th}} = 3.4~\mathrm{V}$ triggers a spike.  
Later, at $t = 14~\mathrm{ns}$ there is an excitatory input the same amplitude which is preceded by an inhibitory signal at $t = 13~\mathrm{ns}$.  
The inhibitory signal prevents the neuron from reaching the firing threshold in response to the excitatory input, and no spike is generated.  
This is an example of negative inhibition, where an inhibitory input suppresses spiking that would otherwise occur. 

Figure~\ref{bio}(d) demonstrates frequency modulation, where a neuron produces a spike train whose frequency depends on the input amplitude.  
When a constant input bias that is above the spike train firing threshold is applied, the NMOS+MTJ neuron generates a continuous series of spikes with a fixed period.  
The spike frequency changes at $t = 11~\mathrm{ns}$ when the input bias amplitude is altered, producing a shorter period between spikes.  
This behavior mirrors the ability of biological neurons to encode stimulus strength in their firing rate.  

Together, the behaviors shown in Figures~\ref{demo} and \ref{bio} represent a subset of biologically inspired dynamics that NMOS+MTJ neurons can intrinsically reproduce using compact, CMOS-compatible circuitry.  
The ability to capture these behaviors in circuit simulations without complex control schemes underscores the potential of MTJ neurons as fundamental building blocks for neuromorphic systems that combine biological realism with practical hardware implementation.

\section{Interconnection of NMOS+MTJ Neurons\label{synapsy}}

Complex computation in both biological and artificial neural networks requires neurons to be interconnected so that activity in one neuron can influence others.  
In biological systems, these connections are formed by synapses, which transmit signals between neurons and determine the timing and strength of neural responses.  
An analogous mechanism is needed in this system to enable information to flow through multiple processing stages.  
The simulations in this section demonstrate that synaptic connections can allow spike propagation across multiple stages, providing a foundation for the machine learning demonstration in the next section.

\begin{figure}
\centering
\includegraphics{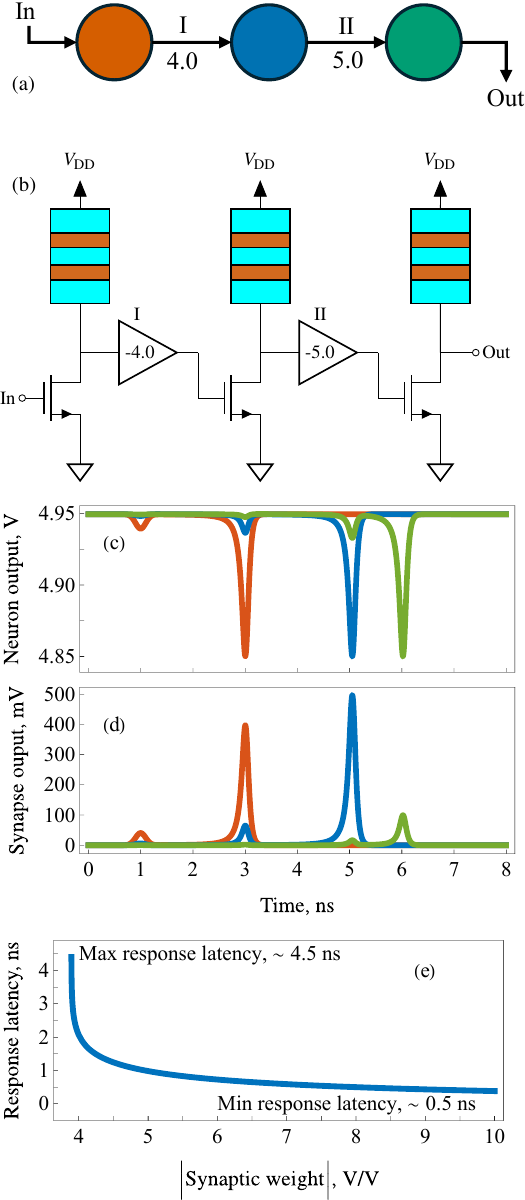}
\caption{
Signal propagation in a chain of NMOS+MTJ neurons with tunable synapses.  
(a) Diagram of a three-neuron chain, where circles represent neurons and arrows represent synapses.  
(b) Electrical schematic of the NMOS+MTJ neuron chain, with each synapse implemented as a tunable voltage amplifier. 
(c) Output voltage waveforms from neuron~1 (orange), neuron~2 (blue), and neuron~3 (green).  
(d) Corresponding synapse outputs, showing the amplifier signals at their respective applied synaptic weights.  
(e) Response latency as a function of synaptic weight, where synaptic weight is expressed as an absolute value.  
}  
\label{network}
\end{figure}

Developing practical synaptic circuits remains an important challenge in neuromorphic hardware.  
The present work does not attempt to solve the broader problem of scalable transistor-level synapse implementation.  
Instead, the synaptic elements are modeled at the circuit-function level in order to study spike propagation, temporal encoding, and learning behavior in NMOS+MTJ neural networks.

Figure~\ref{network}(a) depicts a chain of three neurons, represented by circles, interconnected by two synapses, represented by arrows.  
In the circuit model, each neuron is an NMOS+MTJ pair, as described in the previous section.  
The synapses are modeled as tunable voltage amplifiers that relay signals from one neuron to the next.  
Figure~\ref{network}(b) shows the corresponding electrical circuit schematic for this neuron-synapse chain, where the triangular symbols represent generic amplifiers. 

The circuit was simulated in LTspice using neurons with the same parameters described in Section~\ref{sectmtjnmos}.  
Results are shown in Figures~\ref{network}(c) and~\ref{network}(d).  
Neuron~1, activated by an external input pulse, produces a spike at approximately $t = 3~\mathrm{ns}$, shown in orange in Figure~\ref{network}(c).  
This spike passes through synapse~I, which removes the DC offset and amplifies the signal by $-4~\mathrm{V}/\mathrm{V}$.  
The amplified output, shown in orange in Figure~\ref{network}(d), drives neuron~2, which spikes after a $2~\mathrm{ns}$ latency.  
This output passes through synapse~II, amplified by $-5.0~\mathrm{V}/\mathrm{V}$, and drives neuron~3 after a $1~\mathrm{ns}$ latency.  
The spike from neuron~3, shown in green in Figure~\ref{network}(c), is amplified by $-1~\mathrm{V}/\mathrm{V}$, producing the green waveform in Figure~\ref{network}(d).  
In this simulation, synaptic weight is set by the amplifier gain, with larger gains producing stronger connections and shorter latencies, as seen in the $2~\mathrm{ns}$ latency of neuron~2 (blue) compared to the $1~\mathrm{ns}$ latency of neuron~3 (green).  

This simulation demonstrates that NMOS+MTJ neurons can propagate spiking signals across multiple stages, with the output from each neuron triggering the next.  
The successful transmission of spikes through a chain of connected neurons confirms that these circuits can exhibit both individual neuron-like behavior and coordinated network dynamics.  
This capability suggests a path toward larger neuromorphic hardware systems, where signals propagate through successive processing stages that transform and relay information, analogous to biological neural networks.  

Figure~\ref{network}(e) shows how the synaptic weight, expressed as an absolute value, influences the response latency of the NMOS+MTJ neuron configuration.  
Synaptic weights below approximately $4~\mathrm{V}/\mathrm{V}$ produce a rapid increase in response latency, approaching a maximum of about $4.5~\mathrm{ns}$.  
In contrast, synaptic weights greater than $6~\mathrm{V}/\mathrm{V}$ result in shorter response latencies that converge toward a lower bound of about $0.5~\mathrm{ns}$.  
In this implementation, the synaptic weight is determined by the voltage gain of the amplifier connecting two neurons, and in the simulations that follow we assume that synaptic circuits are capable of supporting a maximum gain of $10~\mathrm{V}/\mathrm{V}$.  

These results demonstrate a tradeoff between timing tunability and timing sensitivity in the NMOS+MTJ neuron system.  
At low synaptic weights, the response latency changes rapidly with small variations in synaptic gain, resulting in increased sensitivity. 
At high synaptic weights, the response latency approaches a saturated minimum value, reducing the ability of the synaptic weight to meaningfully modulate the spike timing.  
The intermediate region of Fig.~\ref{network}(e), therefore, provides a compromise between timing tunability and reduced sensitivity to changes in synaptic gain, and was used for the network simulations presented later in this work. 
For the MTJ and transistor parameters used in this study, this intermediate operating region corresponds to inter-neuron delays typically on the order of $1$--$2~\mathrm{ns}$.  
The representative timing values of approximately 2 and 2.5 ns used later in the network simulations were selected from this region to demonstrate temporal encoding without operating in either the steep near-threshold region or the saturated low-latency region of the simulated response. 

The timing values used in the simulations are not fundamental constraints of the NMOS+MTJ architecture, but are dependent on the MTJ parameters, transistor characteristics, and bias conditions.  
Different MTJ and transistor designs could produce either slower or faster neuron dynamics depending on the intended application.  
In this work, the operating region was chosen as a theoretical demonstration of temporal neuromorphic computation.

The timing region used in the XOR simulations was selected from the intermediate portion of Fig.~\ref{network}(e), where the response latency remains tunable without operating in the highly sensitive near-threshold region or the strongly saturated low-latency region.  
This provides sufficient timing separation for temporal encoding while avoiding large latency shifts from small changes in synaptic weight.  
The effects of timing jitter, device mismatch, and electrical noise were not quantitatively evaluated; therefore, the robustness of the 2 and 2.5 ns encoding to these nonidealities is not established in this work. 

The synapses in this work were simulated using ideal voltage amplifiers in order to study the timing behavior of the NMOS+MTJ neurons. 
Practical implementations could replace these elements with corresponding CMOS circuits such as transconductance amplifiers, programmable gain stages, or circuits using memristive weight storage with fixed-gain transistor stages. 
Because the synaptic and learning circuits are represented by ideal functional blocks, a complete system-level power and energy evaluation was not performed. 
Future work will examine transistor-level synapse designs and their impact on scalability and power consumption.

\section{Machine Learning with an NMOS+MTJ Physical Neural Network\label{sectxor}}

The preceding sections established that NMOS+MTJ neurons reproduce a wide range of biologically inspired behaviors and can propagate spikes through analog synaptic connections. 
These results show that the NMOS+MTJ pair can function as a spiking element within the simulated multineuron network.
We now extend this framework to a trainable spiking neural network modeled using NMOS+MTJ neurons and analog synaptic elements.  
In this architecture, each neuron operates in a single-fire regime, producing at most one spike per input pattern; this timing-based representation makes the neuron firing time differentiable with respect to synaptic weights, enabling gradient-descent learning directly in the analog domain without surrogate gradients or backpropagation-through-time. 
This section presents the network design, the analog signal-flow implementation, and the supervised learning mechanism used to model on-chip training.

To evaluate the learning capability of the NMOS+MTJ network, we apply it to the two-input XOR classification task, a widely used nonlinear test problem in neuromorphic computing and machine learning because it cannot be solved by a single-layer perceptron.
Successful learning of XOR demonstrates nonlinear activation, multilayer processing, and weight adaptation within the LTspice simulation. 
The following subsections describe the network structure, the timing-based gradient-descent rule, and LTspice simulation results showing that the NMOS+MTJ architecture can learn the XOR mapping through analog circuitry.

\begin{figure}
\centering
\includegraphics{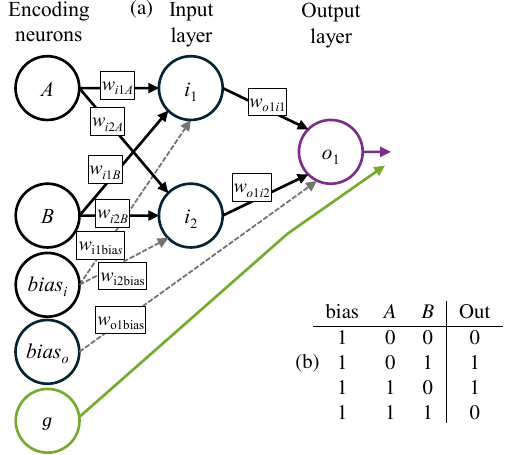}
\caption{
Architecture of the NMOS+MTJ neural network for XOR classification.
(a) Network diagram with encoding neurons, input-layer neurons ($i_1$, $i_2$), and output neuron ($o_1$), and synaptic weights.
(b) XOR truth table with bias input and target output for each input combination.
}
\label{xor}
\end{figure}

Figure~\ref{xor}(a) shows the neural network architecture used for the XOR task.  
The network consists of an input layer with two neurons, $i_1$ and $i_2$, and an output layer with a single neuron, $o_1$.
This configuration represents the minimal structure required to solve the XOR problem, which is not linearly separable.  
The use of multiple layers, nonlinear spiking behavior, and adjustable synaptic weights enables the network to represent the nonlinear decision boundaries needed for correct classification.  

In addition to the primary network layers, external encoding neurons are used to generate the spike patterns that drive the network during both inference and training.  
The system includes five such encoding neurons that operate independently from the neurons within the main layers.  
Neurons $A$ and $B$, shown in Figure~\ref{xor}(a), encode the XOR truth table (Figure~\ref{xor}(b)) inputs as spike waveforms.  
Two additional neurons, $bias_{i}$ and $bias_{o}$, provide constant bias spikes at times chosen specifically for neurons $i_1$, $i_2$, and $o_1$.  
A fifth encoding neuron, named $g$, produces a timed spike that represents the desired output during training and serves as the temporal reference signal for learning.  
Together, these bias and reference spikes implement time encoding within the network, enabling the NMOS+MTJ neural architecture to represent and process information through spike timing.

In the network, neurons communicate through synapses, which are represented by arrows in the diagrams and implemented as voltage amplifiers in the LTspice simulations, as was demonstrated in Section \ref{synapsy}.  
The names of the synapses are labeled in Figure~\ref{xor}(a).  
Because the neurons operate with voltage signals, multiple incoming signals must be combined before reaching the gate of each NMOS+MTJ neuron.  
This combining function is performed by signal adders.  
For example, neuron $i_1$ receives inputs from three separate sources: $A$, $B$, and $bias_{i}$.  
In the LTspice simulation, these inputs are summed using an ideal voltage adder, while practical hardware implementations could employ a standard CMOS summing circuit.  
The summed signal is then applied to the gate of the NMOS+MTJ neuron, where it controls the drain current and thereby influences whether the neuron produces a spike.  

The network uses a fully connected feedforward architecture, where signals move in one direction from the encoding layer to the input layer and then from the input layer to the output layer.  
As shown in Figure~\ref{xor}(a), encoding neurons $A$ and $B$ are connected to both input-layer neurons $i_1$ and $i_2$.  
Each input-layer neuron is connected to the output-layer neuron $o_1$.  
In addition, each of the bias neurons connects to its corresponding target neuron to provide a constant bias input.  

\begin{figure}
\centering
\includegraphics{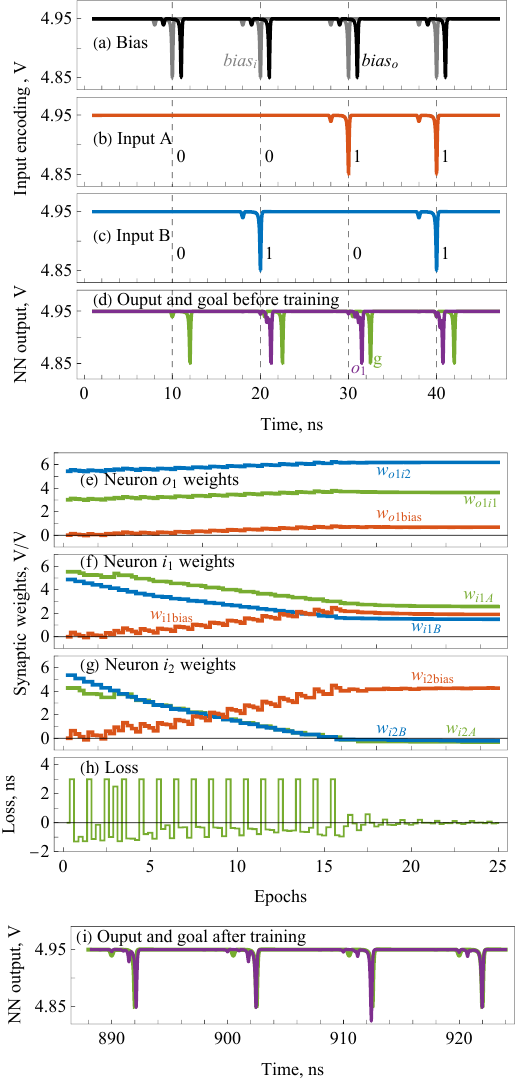}
\caption{
Encoding, inference, and training results for the NMOS+MTJ neural network.  
(a) Bias spike signals for neurons $i_1$ and $i_2$ (gray) and neuron $o_1$ (black).  
(b--c) Spike encoding for inputs $A$ (orange) and $B$ (blue) across the four XOR input conditions: $(0,0)$, $(0,1)$, $(1,0)$, and $(1,1)$.  
(d) Before training, output spikes from neuron $o_1$ (purple) do not align with target spikes from neuron $g$ (green).  
(e--g) Evolution of individual synaptic weights over training epochs, showing convergence by approximately epoch~16.  
(h) Loss function over training epochs, with a sharp drop around epoch~16 indicating successful learning.  
(i) After training, output spikes from neuron $o_1$ (purple) align with target spikes from neuron $g$ (green).  
}
\label{xorcycle}
\end{figure}

Inference for the XOR truth table is performed by sequentially presenting input spikes, with each row encoded at $10~\mathrm{ns}$ intervals.  
A full inference cycle, or epoch, takes approximately $40~\mathrm{ns}$, corresponding to the four rows of the XOR truth table.  
Figures~\ref{xorcycle}(a) through \ref{xorcycle}(c) illustrate how each row of the truth table is encoded.  
Figure~\ref{xorcycle}(a) shows the spikes from the bias neuron $bias_{i}$, plotted in gray, occurring at $t = 10$, $20$, $30$, and $40~\mathrm{ns}$, once for each truth table row.  
The bias neuron $bias_{o}$, plotted in black, fires $1~\mathrm{ns}$ later than $bias_{i}$ because it is located in the downstream output layer.  

Figures~\ref{xorcycle}(b) and \ref{xorcycle}(c) show the spikes generated by neurons $A$ and $B$ for each row of the XOR truth table.  
At $t = 10~\mathrm{ns}$, corresponding to inputs $A = 0$ and $B = 0$, neither neuron $A$ nor neuron $B$ fires, and only the bias neurons produce spikes.  
At $t = 20~\mathrm{ns}$, corresponding to inputs $A = 0$ and $B = 1$, neuron $B$ fires together with the bias neurons.  
At $t = 30~\mathrm{ns}$, corresponding to inputs $A = 1$ and $B = 0$, neuron $A$ fires together with the bias neurons.  
At $t = 40~\mathrm{ns}$, corresponding to inputs $A = 1$ and $B = 1$, both neuron $A$ and neuron $B$ fire together with the bias neurons.  

Figure~\ref{xorcycle}(d) shows the output response of the neural network before training.  
The expected spike times, referred to as the ``goal'' times, are shown in green and are generated by neuron $g$.  
In the XOR encoding scheme used here, a spike occurring $2.0~\mathrm{ns}$ after the bias input represents a logical zero, while a spike occurring $2.5~\mathrm{ns}$ after the bias input represents a logical one. 
These timing values were selected from the intermediate operating region identified in Fig.~\ref{network}(e), where the response latency remains sufficiently tunable, while avoiding both the highly sensitive near-threshold regime and the strongly saturated low-latency regime.  
The actual output spikes from neuron $o_1$ are shown in purple and result from the propagation of the encoded input signals through the network.  
Before training, all synaptic weights were initialized to random values between $3$ and $6~\mathrm{V}/\mathrm{V}$, with the exception of the bias weights, which were set to zero.  
As shown in Figure~\ref{xorcycle}(d), the untrained network fails to produce a spike for the first input condition, and the spikes generated for the remaining cases occur at incorrect times.  
This mismatch between the output and the goal times demonstrates the need for training to align the network behavior with the desired XOR function.  

To achieve correct performance, the network must adjust the synaptic weights so that the output neuron $o_1$ fires at the goal spike times, expressed in nanoseconds, defined by the reference neuron $g$.  
This alignment is achieved using a supervised learning strategy that follows the general structure of feedforward operation and gradient-based error backpropagation \cite{kheradpisheh2020temporal}, but with modifications tailored to our analog architecture. 
Each synaptic weight is denoted by $w_{k,j}$, where $j$ is the presynaptic (upstream) neuron and $k$ is the postsynaptic (downstream) neuron.  
The update to each weight, $\Delta w_{k,j}$, follows the rule  
\begin{equation}
\Delta w_{kj} = -\eta \frac{dL}{dt_k} \frac{\partial t_k}{\partial w_{k,j}}\label{upsy}
\end{equation}  
where $\eta$ is the learning rate, $L$ is the loss function, and $t_j$ is the firing time of the presynaptic neuron.  
The loss function is defined as  
\begin{equation}
L = \frac{1}{2} (t_{\mathrm{actual}} - t_{\mathrm{desired}})^2
\end{equation}  
with $t_{\text{actual}}$ representing the firing time of neuron $o_1$ and $t_{\text{desired}}$ the target spike time specified by neuron $g$.  
The derivative of the loss function with respect to firing time is  
\begin{equation}
\frac{dL}{dt_j} = \frac{\partial L}{\partial t_j} + \sum_{k<j} \frac{dL}{dt_k} \cdot \frac{\partial t_k}{\partial t_j}
\end{equation}  
which accounts for both direct and indirect timing dependencies within the network.  

Completing the gradient update in (\ref{upsy}) requires evaluating the terms $\partial t_k/\partial w_{k,j}$ and $\partial t_k/\partial t_j$, which quantify how the firing time of the postsynaptic neuron changes in response to variations in the synaptic weight or the spike time of the presynaptic neuron, respectively.  
In this work, these derivatives are approximated as  
\begin{equation}
\frac{\partial t_k}{\partial w_{k,j}} = u(t_k - t_j)
\end{equation}  
\begin{equation}
\frac{\partial t_k}{\partial t_j} = w_{k,j} \, u(t_k - t_j)
\end{equation}  
where $u(~)$ is the unit step function, which enforces causality by ensuring that only presynaptic spikes occurring before the postsynaptic spike can contribute to weight changes that advance or delay the postsynaptic firing time.  
Additional details on how the gradient descent learning rule was implemented in analog form are provided in Appendix~B.  

The network was trained entirely in LTspice using repeated cycles of inference, with each cycle followed by synaptic weight adjustment based on the gradient descent learning rule described earlier.  
The four input patterns of the XOR truth table, $(0,0)$, $(0,1)$, $(1,0)$, and $(1,1)$, were applied sequentially during each training epoch, giving a total epoch duration of $40~\mathrm{ns}$.  
Each pattern occupied a fixed $10~\mathrm{ns}$ interval.  
Within each interval, the network first executed feedforward inference to generate a spike response at the output neuron.  
Because the NMOS+MTJ neuron exhibits a typical response latency of less than $5~\mathrm{ns}$, the remaining portion of the interval between approximately $5$ and $10~\mathrm{ns}$ was available for error evaluation and synaptic weight updates.  
The timing error between the observed and target spike times was computed, and the gradient descent rule was applied to adjust the synaptic weights.  
This procedure was repeated for each input pattern in succession, completing one full training epoch with four weight updates.  

Figures~\ref{xorcycle}(e)--\ref{xorcycle}(g) show the evolution of individual synaptic weights, which gradually converged toward stable values at the sixteenth epoch.  
As shown in Figure~\ref{xorcycle}(h), the loss function remained high during the early epochs, dropped sharply after epoch 16, and approached zero thereafter, indicating successful alignment between the network output and the target spike times.  
The weight convergence and loss minimization are consistent with the NMOS+MTJ neural network having learned the XOR mapping in the simulated hardware environment.  

Once the network converges, the output spikes from neuron $o_1$ align with the target spikes from neuron $g$ for all input conditions.  
This alignment is illustrated for epoch 22 occurring at approximately $t = 900~\mathrm{ns}$ in Figure~\ref{xorcycle}(i). 
The figure shows that the output spikes from the neural network (purple) match closely with the spikes from neuron $g$ (green), confirming that the network successfully learned the XOR function.  

Tables~\ref{wwwwaits} and~\ref{ttthymes} report the trained weights and final timing results at epoch 22. 
Training was continued through epoch 22 to further reduce the residual timing error. At epoch 22, the maximum absolute timing error across the four XOR input conditions was 0.102 ns.

\begin{table}[ht]
	\centering
		\caption{Trained weights at epoch 22}
		\vspace{-0.5mm} 
		\label{wwwwaits}
		\begin{tabular}{cc}
			\hline \vspace{.3mm}
			\textbf{Synaptic weight} & \textbf{Final value (V/V)} \\
			\hline 
			$w_{o1i1}$ & $3.63$ \\
			$w_{o1i2}$ & $6.19$ \\
			$w_{i1A}$ & $2.56$ \\
			$w_{i1B}$ & $1.49$ \\
			$w_{i2A}$ & $-0.33$ \\
			$w_{i2B}$ & $-0.22$ \\
			$w_{\mathrm{bias}_{o1}}$ &  $0.68$ \\
			$w_{\mathrm{bias}_{i1}}$ &  $1.90$ \\
			$w_{\mathrm{bias}_{i2}}$ &  $4.26$  \vspace{1mm}\\
			\hline
		\end{tabular}
\end{table}

\begin{table}[ht]
	\centering
		\caption{Final XOR timing results at epoch 22}
		\vspace{-0.5mm}
		\label{ttthymes}
		\begin{tabular}{cccc}
			\hline \vspace{.3mm}
			\textbf{Input} & \textbf{Target delay (ns)} & \textbf{Output delay (ns)} & \textbf{Timing error (ns)} \\
			\hline
			0, 0 & 2.005 & 2.107 & 0.102 \\
			0, 1 & 2.499 & 2.458 & -0.041 \\
			1, 0 & 2.503 & 2.413 & -0.090 \\
			1, 1 & 2.005 & 1.983 & -0.022 \\
			\hline
		\end{tabular}
\end{table}

It is important to note that the entire training process for the XOR classification task was performed within the LTspice simulation environment using only analog circuitry.  
Both the neuron dynamics and the learning algorithm, including gradient descent and spike-timing-based weight updates, were designed using analog circuit components without external data processing.  
This provides a demonstration that machine learning for a nonlinear classification problem can be modeled in the analog domain using time-encoded spike signals.
  
By demonstrating end-to-end supervised learning within LTspice, this work provides a proof-of-concept for a CMOS+X simulation that combines training and inference within a single platform.
The NMOS+MTJ neuron illustrates this concept by modeling the dynamic operation of an MRAM-derived 1T-1MTJ structure within the simulated network.

The network learned the XOR function through continuous cycles of spike generation, inference, error evaluation, and synaptic weight adjustment, all governed by voltage-based interactions in the simulated circuit.  
The total simulated training time was approximately $1~\mu\mathrm{s}$, and the complete LTspice simulation can be completed in about $15$ seconds on a conventional laptop.
These results provide an example of simulated trainable neuromorphic computation using spintronic device models and analog circuit elements.

\section{Conclusion}
This work has presented the design, simulation, and evaluation of an NMOS+MTJ neuron architecture capable of reproducing multiple biologically inspired behaviors and supporting machine learning through gradient-based weight adaptation.  
Starting from the demonstration of single-neuron properties such as threshold activation, spike generation, response latency, and synaptic integration, the study progressed to multi-neuron configurations that successfully propagated spike signals through tunable synapses.  
Extending these capabilities, a three-layer NMOS+MTJ neural network was constructed and trained in LTspice to perform the two-input XOR task, where it achieved convergence under modeled conditions. 
Both neuron dynamics and the learning algorithm were modeled using analog components without digital computation, showing that time-encoded spike signals and analog circuitry can support integrated inference and learning within LTspice analog circuit simulations.   
These results motivate further investigation of NMOS+MTJ neurons as potentially scalable building blocks for neuromorphic systems that could link nanoscale spintronic devices with CMOS-compatible analog electronics.
Experimental validation, transistor-level implementation of the ideal circuit elements, quantitative power evaluation, robustness testing, and extension to larger networks remain subjects for future work.  

\vspace{-2mm}
\section*{Acknowledgments}
I.N.K., M.B.A., and G.D.N. acknowledge partial support from the National Science Foundation via awards ECCS-2213690 and DMREF-2324203 as well as from the University of California National Laboratory Fees Research Program.

\section*{Appendix A: Simulation Model and Parameters}

Accurate simulation of the NMOS+MTJ neuron in a standard electrical circuit simulator is important for validating device behavior, exploring circuit-level interactions, and ensuring compatibility with conventional CMOS circuitry.  
MTJs can be simulated by equivalent circuit models, as in Figures~1(b) and~1(c) of~\cite{phyBased}.  
These models allow the magnetization dynamics of the MTJ free layer to be solved alongside electronic circuit elements in a unified SPICE environment, enabling both device-level and system-level analysis.

\begin{figure}
\centering
\includegraphics{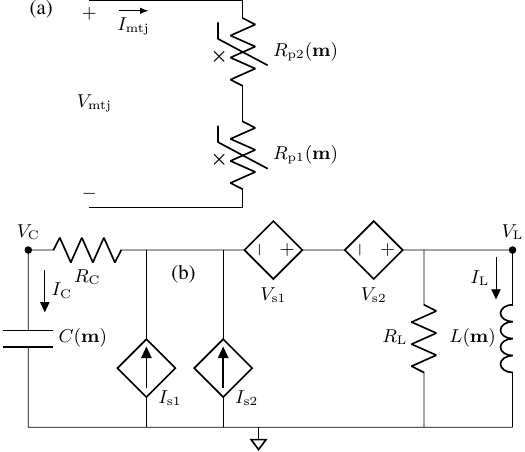}
\caption{Circuit models for the MTJ.
(a) Equivalent circuit model of the three-layer MTJ, represented as two intrinsic magnetoresistors in series: $R_{p1}(\phi)$ for the reference and free layers, and $R_{p2}(\phi)$ for the analyzer and free layers.  
(b) Spin-circuit representation used to determine the free layer magnetization orientation $\mathbf{m}$ from the calculated charge current $I_C$ and voltage $V_L$.  
The spin-circuit is based on~\cite{phyBased} with additional dependent sources.  
}
\label{spice}
\end{figure}

In this work, we extend the two-ferromagnetic-layer MTJ model of~\cite{phyBased} to represent an MTJ with three ferromagnetic layers, as depicted in Figure~\ref{toon}(a).  
The corresponding electrical equivalent, shown in Figure~\ref{spice}(a), models the total MTJ resistance as two magnetoresistive elements connected in series:  
\begin{equation}
R_\mathrm{MTJ}(\phi) = R_{p1}(\phi) + R_{p2}(\phi) ,
\label{allRs}
\end{equation}
where $R_{p1}(\phi)$ represents the resistance from $\mathbf{p}_1$ and $\mathbf{m}$, while $R_{p2}(\phi)$ represents the resistance from $\mathbf{p}_2$ and $\mathbf{m}$.

Both $R_{p1}(\phi)$ and $R_{p2}(\phi)$ vary explicitly with the free layer magnetization angle $\phi$, making accurate determination of $\phi$ important to reproducing the MTJ response.  
The angle $\phi$ is obtained by solving the spin-circuit model shown in Figure~\ref{spice}(b), which calculates $\mathbf{m}$ from the charge current $I_C$ and voltage $V_L$.  
This spin-circuit is adapted from~\cite{phyBased} and includes additional dependent sources to capture the behavior of the extended three-layer structure.

All modeling elements were implemented in LTspice, enabling the MTJ to be simulated alongside CMOS-compatible electronic components such as NMOS transistors, voltage sources, and amplifiers.  
The NMOS transistor is taken from the default LTspice component library, and is connected in series with the MTJ to form the neuron core.  
The positive supply voltage $V_\mathrm{DD}$ is fixed at $5.0~\mathrm{V}$ and serves as the drain supply for the NMOS.  

The simulation parameters are summarized in Table~\ref{params}.  
The MTJ free layer is modeled as an ellipse with semiaxes of $30~\mathrm{nm}$ and $20~\mathrm{nm}$, and a thickness of $3~\mathrm{nm}$.  
Interface resistances for the parallel ($R_\mathrm{P}$) and antiparallel ($R_\mathrm{AP}$) states are specified independently for both the reference-free and analyzer-free layer pairs.  
The polar angle $\theta$ and azimuthal angle $\varphi$ define the orientation of each magnetic layer.

\begin{table}[ht]
\begin{center}
\caption{NMOS+MTJ simulation parameters.}
\vspace{-1mm} 
\label{params}
\begin{tabular}{ll}
\hline
\textbf{Parameter} & \textbf{Value} \\
\hline
Free layer volume ($\mathrm{nm^3}$) & $5654.87$ \\
Saturation magnetization $M_s$ (A/m) & $795775$ \\
Gyromagnetic ratio $\gamma$ (GHz/T) & $2\pi 28$ \\
Gilbert damping $\alpha_G$ & $0.1$ \\
External field $B_e$ (mT) & $5.3$ \\
Demagnetizing field $B_d$, (T) & 1 \\
Anisotropy field $B_a$ (T) & $0$ \\
Analyzer layer orientation $(\theta,~\varphi)$ & $(\pi/2,~\pi/2)$ \\
Reference layer orientation $(\theta,~\varphi)$ & $(0,~0)$ \\
Free layer initial orientation $(\theta,~\varphi)$ & $(\pi/2,~0)$ \\
Analyzer-free layer resistances $R_\mathrm{P,p2},R_\mathrm{AP,p2}$ ($\Omega$) & $500,~1500$ \\
Reference-free layer resistances $R_\mathrm{P,p1},R_\mathrm{AP,p1}$ ($\Omega$) & $0.5,~1.5$ \\
Drain supply $V_\mathrm{DD}$ (V) & $5.0$ \\
\hline
\end{tabular}
\end{center}
\end{table}

\section*{Appendix B: Analog Circuits for Gradient-Descent Learning}

In the main text, the gradient descent learning algorithm was implemented entirely within the LTspice simulation environment using analog circuit components.  
Section~\ref{sectxor} presented the derivation of the learning rule, beginning with the loss function and its derivatives, and showed how spike timing differences drive synaptic weight updates in the NMOS+MTJ neural network.  
This appendix provides details about performing these operations in analog circuit form.   

The learning rule in Section~\ref{sectxor} was expressed in general form using gradient descent with spike-timing-dependent derivatives.  
Here, we restate the explicit implementation used in our LTspice simulations.  
These equations apply to all synapses in the network, including bias connections.  
The complete set of update equations is  
\begin{equation}
\begin{aligned}
\Delta w_{o1i1} &= -\eta d_o u(t_{o1} - t_{i1}), \\
\Delta w_{o1i2} &= -\eta d_o u(t_{o1} - t_{i2}), \\
\Delta w_{i1A} &= -\eta d_o w_{o1,i1} a(t) u(t_{o1} - t_{i1}), \\
\Delta w_{i1B} &= -\eta d_o w_{o1,i1} b(t) u(t_{o1} - t_{i1}), \\
\Delta w_{i2A} &= -\eta d_o w_{o1,i2} a(t) u(t_{o1} - t_{i2}), \\
\Delta w_{i2B} &= -\eta d_o w_{o1,i2} b(t) u(t_{o1} - t_{i2}), \\
\Delta w_{\mathrm{bias}_{o1}} &= -\eta d_o, \\
\Delta w_{\mathrm{bias}_{i1}} &= -\eta d_o w_{o1i1}, \\
\Delta w_{\mathrm{bias}_{i2}} &= -\eta d_o w_{o1i2}.
\end{aligned}
\label{explic}
\end{equation}
Here, $d_o = t_{\mathrm{actual}} - t_{\mathrm{desired}}$ is the time difference between the spike from output neuron $o_1$ and the target spike from reference neuron $g$.  
The functions $a(t)$ and $b(t)$ represent the binary inputs `0' and `1' for $A$ and $B$ in the truth table. 

\begin{figure}
\centering
\includegraphics{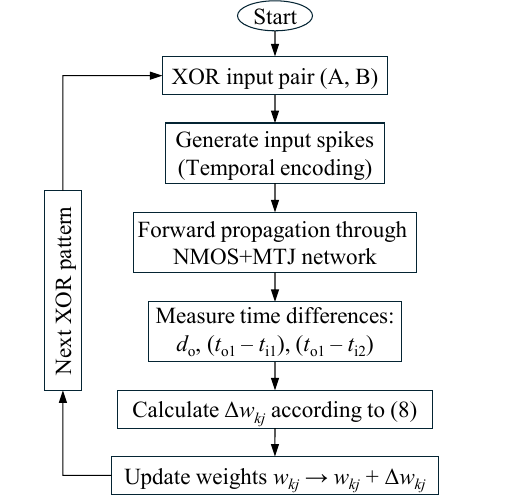}
\caption{Flow chart of the XOR training sequence used to generate input spikes, propagate them through the NMOS+MTJ network, compute timing-based weight updates, and advance to the next training pattern. }\label{flo}
\end{figure}

Figure~\ref{flo} summarizes the training sequence used for each XOR input pattern.  
For each pattern, the corresponding input spikes are generated using the encoding scheme described in Section~\ref{sectxor}.  
The spikes, then, propagate through the NMOS+MTJ network, producing output spikes with different delay times at the downstream neurons.  
The learning circuit measures the timing differences $d_o$ and the relative spike times $(t_{o1}-t_{i1})$ and $(t_{o1}-t_{i2})$.  
These timing quantities are then used to calculate the synaptic weight updates $\Delta w_{kj}$, after which the stored weights are updated before the next XOR input pattern is applied.

From (\ref{explic}), it follows that only three functional blocks are required to implement the weight update rule in an analog circuit simulation: (i) subtraction to compute $d_o$ and other spike timing differences, (ii) multiplication to calculate the weight changes, and (iii) an analog register to store the synaptic weight. 
The timing-difference operation was implemented with a gated current source charging a capacitor, so that the stored voltage is proportional to the time interval between two spike events.
The multiplication operations in (\ref{explic}) were implemented using LTspice behavioral sources that directly compute the required products of timing-error signals, synaptic weights, and input terms.
The synaptic weight $w_{kj}$ was represented as a voltage, and stored using a sample-and-hold block, allowing the value to be retained between training updates and updated after each input pattern.

Although behavioral models were used for expedience in simulation, each function can, in principle, be realized using practical CMOS-compatible analog circuitry.  
The gain values that define the synaptic weights could be stored in an analog register implemented with a capacitor to hold a voltage.  
These stored voltages would directly control the tunable amplifiers that implement the synapses, ensuring that each amplifier maintains its programmed gain between training updates.  
Spike timing detection could be achieved in the analog domain using monostable circuits or pulse-width measurement blocks that generate a voltage proportional to the time interval between two events.  
In this application, such a circuit would measure the time difference between pre-synaptic and post-synaptic spikes directly as an analog voltage, providing the $d_o$ term in (\ref{explic}) without any conversion to the digital domain.  
Multiplication could be performed with a simple analog scheme, such as a single-quadrant multiplier using a MOSFET in the triode region, where the drain current is proportional to the product of two input voltages.  
This approach would complete the weight update computation while remaining compatible with standard CMOS processes.  
These examples are presented as illustrative pathways for eventual hardware implementation and are consistent with the nature of this work, in which the emphasis is on demonstrating functional feasibility rather than delivering a fabrication-ready design.  

\vfill


\vspace{-2mm}
\begin{thebibliography}{1}

\bibitem{li2019edge}
E.~Li, L.~Zeng, Z.~Zhou, and X.~Chen, ``Edge AI: On-demand accelerating deep neural network inference via edge computing,'' \textit{IEEE Trans. Wireless Commun.}, vol.~19, no.~1, pp. 447--457, 2020.

\bibitem{jouppi2017datacenter}
N.~P. Jouppi, C.~Young, N.~Patil, D.~Patterson, G.~Agrawal, R.~Bajwa, \textit{et~al.}, ``In-datacenter performance analysis of a tensor processing unit,'' in \textit{Proc. 44th Annu. Int. Symp. Comput. Archit. (ISCA)}, 2017, pp. 1--12.

\bibitem{maass1997networks}
W.~Maass, ``Networks of spiking neurons: the third generation of neural network models,'' \textit{Neural Netw.}, vol.~10, no.~9, pp. 1659--1671, 1997.

\bibitem{davies2018loihi}
M.~Davies, N.~Srinivasa, T.-H. Lin, G.~Chinya, Y.~Cao, S.~H. Choday, \textit{et~al.}, ``Loihi: A neuromorphic manycore processor with on-chip learning,'' \textit{IEEE Micro}, vol.~38, no.~1, pp. 82--99, 2018.

\bibitem{markovic2020physics}
D. Markovi\'c, A. Mizrahi, D. Querlioz, and J. Grollier, ``Physics for neuromorphic computing,'' \textit{Nat. Rev. Phys.}, v. 2, no. 9, pp. 499-510, 2020.

\bibitem{tang2023seneca}
G.~Tang, K.~Vadivel, Y.~Xu, R.~Bilgic, K.~Shidqi, P.~Detterer, \textit{et al.}, ``SENECA: building a fully digital neuromorphic processor, design trade-offs and challenges,'' \textit{Front. Neurosci.}, vol.~17, p.~1187252, 2023.


\bibitem{grollier2016bioinspired}
J.~Grollier, D.~Querlioz, and M.~D.~Stiles, ``Spintronic nanodevices for bioinspired computing,'' \textit{Proceedings of the IEEE}, vol.~104, no.~10, pp.~2024--2039, 2016.

\bibitem{finocchio2024roadmap}
G.~Finocchio, J.~Incorvia, J.~Friedman, Q. Yang, Qu, A. Giordano, J. Grollier, \textit{et al.}, ``Roadmap for unconventional computing with nanotechnology,'' \textit{Nano Futures}, vol.~8, no.~1, p.~012001, 2024.

\bibitem{Incorvia}
J. Incorvia, P. Xiao, N. Zogbi, A. Naeemi, C. Adelmann, F. Catthoor, \textit{et al.}, ``Spintronics for achieving system-level energy-efficient logic,'' \textit{Nat Rev Electr Eng}, vol. 1, no. 11, pp. 700-713, Oct. 2024.

\bibitem{indiveri2015memory}
G.~Indiveri and S.~Liu, ``Memory and information processing in neuromorphic systems,'' \textit{Proc. IEEE}, vol.~103, no.~8, pp.~1379--1397, 2015.

\bibitem{nsf2022cmosx}
National Science Foundation, ``Partnership for Prototyping of CMOS+X Systems,'' \textit{Dear Colleague Letter NSF-22-076}, Apr. 22, 2022. [Online]. Available: https://www.nsf.gov/funding/opportunities/dcl-partnership-prototyping-cmosx-systems.  

\bibitem{chappert}
C. Chappert, A. Fert, and F. Van Dau, ``The emergence of spin electronics in data storage,'' \textit{Nature Mater}, vol. 6, no. 11, pp. 813-823, 2007.

\bibitem{alzate}
J. G. Alzate, P. K. Amiri, and K. L. Wang, ``Magnetic Tunnel Junctions and Their Applications in Non-volatile Circuits,'' \textit{Handbook of Spintronics. Springer Netherlands}, pp. 1127-1171, 2016.

\bibitem{edelstein}
D.~Edelstein, M.~Rizzolo, D.~Sil, A.~Dutta, J.~DeBrosse, M.~Wordeman, \textit{et al.}, ``A 14 nm embedded STT-MRAM CMOS technology,'' in \textit{Proc. IEEE Int. Electron Devices Meeting (IEDM)}, 2020, pp.~11--5.


\bibitem{everspin}
Everspin Technologies, Inc., ``Everspin Technologies,'' Accessed: Aug. 11, 2025. [Online]. Available: \url{https://www.everspin.com/}

\bibitem{sun}
J. Z. Sun and D. C. Ralph, ``Magnetoresistance and spin-transfer torque in magnetic tunnel junctions,'' \textit{Journal of Magnetism and Magnetic Materials}, vol. 320, no. 7, pp. 1227-1237, Apr. 2008.

\bibitem{mtjneuron}
S. Louis, H. Bradley, C. Trevillian, A. Slavin, and V. Tyberkevych, ``Spintronic Neuron Using a Magnetic Tunnel Junction for Low-Power Neuromorphic Computing,'' \textit{IEEE Magn. Lett.}, vol. 15, pp. 1-5, 2024.

\bibitem{salinas2023lastmile}
R.~Salinas, P.~Chen, C.~Yang, and C.~Lai, ``Spintronic materials and devices towards an artificial neural network: accomplishments and the last mile,'' \textit{Materials Research Letters}, v.11, no.5, 305--326, 2023.

\bibitem{chumak2022roadmap}
A.~V.~Chumak, P.~Kabos, M.~Wu, C.~Abert, C.~Adelmann, A.~O.~Adeyeye, \textit{et al.}, ``Advances in magnetics roadmap on spin-wave computing,'' \textit{IEEE Trans. Magn.}, vol.~58, no.~6, pp.~1--72, 2022.


\bibitem{jiang2024spin}
S. Jiang, L. Yao, S. Wang, D. Wang, L. Liu, \textit{et al.}, ``Spin-torque nano-oscillators and their applications,'' \textit{Appl. Phys. Rev.}, vol. 11, no. 4, 2024.

\bibitem{Lin2009}
C.~J.~Lin, S.~H.~Kang, Y.~J.~Wang, K.~Lee, X.~Zhu, W.~C.~Chen, \textit{et al.}, ``45nm low power CMOS logic compatible embedded STT MRAM utilizing a reverse-connection 1T/1MTJ cell,'' in \textit{Proc. IEEE Int. Electron Devices Meeting (IEDM)}, pp.~1-4, 2009.  

\bibitem{phyBased}
S. Louis, H. Bradley, A. Litvinenko, and V. Tyberkevych, ``A Physics-Based Circuit Model for Magnetic Tunnel Junctions,'' \textit{IEEE Magn. Lett.}, vol. 16, pp. 1-5, 2025, doi: 10.1109/lmag.2025.3577475.

\bibitem{slon1}
J. C. Slonczewski, ``Conductance and exchange coupling of two ferromagnets separated by a tunneling barrier,'' \textit{Phys. Rev. B}, vol.39, no. 10, pp. 6995-7002, Apr. 1989.

\bibitem{slon2}
J. C. Slonczewski and J. Z. Sun, ``Theory of voltage-driven current and torque in magnetic tunnel junctions,'' \textit{Journal of Magnetism and Magnetic Materials}, vol. 310, no. 2, pp. 169-175, Mar. 2007.

\bibitem{krivorotov2008time}
I.~N.~Krivorotov, N.~C.~Emley, R.~A.~Buhrman, and D.~C.~Ralph, ``Time-domain studies of very-large-angle magnetization dynamics excited by spin transfer torques,'' \textit{Phys. Rev. B}, vol.~77, no.~5, p.~054440, Feb.~2008.

\bibitem{worledge}
D.~Worledge, G.~Hu, D.~Abraham, J.~Sun, P.~Trouilloud, J.~Nowak, \textit{et al.}, ``Spin torque switching of perpendicular {TaCoFeBMgO}-based magnetic tunnel junctions,'' \textit{Appl. Phys. Lett.}, vol.~98, no.~2, p.~022501, 2011.


\bibitem{Bhatti2017}
S.~Bhatti, R.~Sbiaa, A.~Hirohata, H.~Ohno, S.~Fukami, and S.~N.~Piramanayagam, ``Spintronics based random access memory: A review,'' \textit{Materials Today}, vol.~20, no.~9, pp.~530–548, 2017. 

\bibitem{Nguyen2024}
V.~D.~Nguyen, S.~Rao, K.~Wostyn, and S.~Couet, ``Recent progress in spin-orbit torque magnetic random-access memory,'' \textit{npj Spintronics}, vol.~2, article~48, 2024.

\bibitem{torrejon2017}
J.~Torrejon, M.~Riou, F.~A.~Araujo, S.~Tsunegi, G.~Khalsa, D.~Querlioz, \textit{et al.}, ``Neuromorphic computing with nanoscale spintronic oscillators,'' \textit{Nature}, vol.~547, pp.~428--431, 2017.

\bibitem{riou2019}
M.~Riou, J.~Torrejon, B.~Garitaine, F.~Abreu~Araujo, P.~Bortolotti, V.~Cros, \textit{et al.}, ``Temporal pattern recognition with delayed-feedback spin-torque nano-oscillators,'' \textit{Phys. Rev. Appl.}, vol.~12, p.~024049, 2019.

\bibitem{romera2018}
M.~Romera, P.~Talatchian, S.~Tsunegi, F.~A.~Araujo, V.~Cros, P.~Bortolotti, \textit{et al.}, ``Vowel recognition with four coupled spin-torque nano-oscillators,'' \textit{Nature}, vol.~563, pp.~230--234, 2018.

\bibitem{Zhou2021}
P.~Zhou, A.~Edwards, F.~Mancoff, D.~Houssameddine, S.~Aggarwal, and J.~Friedman, ``Experimental demonstration of neuromorphic network with STT-MTJ synapses,''  \textit{arXiv preprint arXiv:2112.04749}, 2021.  

\bibitem{kaiser2022hardware}
J.~Kaiser, W.~Borders, K.~Camsari, S.~Fukami, H.~Ohno, and S.~Datta, ``Hardware-aware in situ learning based on stochastic magnetic tunnel junctions,'' \textit{Phys. Rev. Appl.}, vol.~17, no.~1, p.~014016, Jan.~2022.

\bibitem{Goodwill2022}
J.~Goodwill, N.~Prasad, B.~Hoskins, M.~Daniels, \textit{et al.}, ``Implementation of a binary neural network on a passive array of magnetic tunnel junctions,'' \textit{Phys. Rev. Appl.}, vol.~18, no.~1, p.~014039, 2022.

\bibitem{Rzeszut2022}
P.~Rzeszut, J.~Checinski, I.~Brzozowski, S.~Zietek, W.~Skowronski, and T.~Stobiecki, ``Multi-state MRAM cells for hardware neuromorphic computing,'' \textit{Scientific Reports}, vol.~12, article~7178, 2022. 

\bibitem{you20238b}
D. You, Y.~Chiu, W.~Khwa, C.~Li, F.~Hsieh, \textit{et al.}, ``An 8b-precision 8-Mb STT-MRAM near-memory-compute macro using weight-feature and input-sparsity aware schemes for energy-efficient edge AI devices,'' \textit{IEEE Journal of Solid-State Circuits}, vol.~59, no.~1, pp.~219--230, 2023.

\bibitem{Lv2024}
Y.~Lv, B.~Zink, R.~Bloom, H.~Cılasun, P.~Khanal, \textit{et al.}, ``Experimental demonstration of magnetic tunnel junction-based computational random-access memory,'' \textit{npj Unconventional Computing}, vol.~1, article~3, 2024. 

\bibitem{Kiseleva2025}
K.~Kiseleva, D.~Cherkasov, G.~Kichin, V.~Antonov, and K.~Zvezdin, ``Influence of parameters of crossbars with STT-MRAM on the accuracy of analog neural network,'' \textit{Bulletin of the Russian Academy of Sciences: Physics}, vol.~89, no.~5, pp.~774–779, 2025. 

\bibitem{Lee2025}
J.~Lee, J.~Song, H.~Im, J.~Kim, W.~Lee, W.~Yi, \textit{et al.}, ``A scalable neural network emulator with MRAM-based mixed-signal circuits,'' \textit{Frontiers in Neuroscience}, vol.~19, article~1599144, 2025.

\bibitem{Khymyn}
R.~Khymyn, I.~Lisenkov, J.~Voorheis, O.~Sulymenko, O.~Prokopenko, V.~Tiberkevich, \textit{et al.}, ``Ultra-fast artificial neuron: generation of picosecond-duration spikes in a current-driven antiferromagnetic auto-oscillator,'' \textit{Sci. Rep.}, vol.~8, no.~1, p.~15727, 2018.


\bibitem{Liu}
Y. Liu, I. Barsukov, Y. Barlas, I. N. Krivorotov, and R. K. Lake, ``Synthetic antiferromagnet-based spin Josephson oscillator,'' \textit{Applied Physics Letters}, vol. 116, no. 13, Mar. 2020. 

\bibitem{Mitro}
A. Mitrofanova, A. Safin, O. Kravchenko, S. Nikitov, and A. Kirilyuk, ``Optically initialized and current-controlled logical element based on antiferromagnetic-heavy metal heterostructures for neuromorphic computing,'' \textit{Applied Physics Letters}, vol. 120, no. 7, Feb. 2022.

\bibitem{Ovcha}
R. V. Ovcharov, E. G. Galkina, B. A. Ivanov, and R. S. Khymyn, ``Spin Hall Nano-Oscillator Based on an Antiferromagnetic Domain Wall,'' \textit{Phys. Rev. Applied}, vol. 18, no. 2, Aug. 2022.

\bibitem{Crotty2010}
P. Crotty, D. Schult, and K. Segall, ``Josephson junction simulation of neurons,'' \textit{Phys. Rev. E}, vol. 82, no. 1, July 2010.

\bibitem{Schneid}
M. L. Schneider, C. A. Donnelly, and S. E. Russek, ``Tutorial: High-speed low-power neuromorphic systems based on magnetic Josephson junctions,'' \textit{Journal of Applied Physics}, vol. 124, no. 16, Oct. 2018. 

\bibitem{rodrigues2023spintronic}
D.~R.~Rodrigues, R.~Moukhader, Y.~Luo, B.~Fang, A.~Pontlevy, A.~Hamadeh, Z.~Zeng, M.~Carpentieri, and G.~Finocchio, ``Spintronic Hodgkin--Huxley--analogue neuron implemented with a single magnetic tunnel junction,'' \textit{Phys. Rev. Appl.}, vol.~19, no.~6, p.~064010, June 2023.

\bibitem{zhang2025dual}
L.~Zhang, Z.~Liu, Y.~Wang, S.~Liu, W.~Wang, B.~Fang, \textit{et al.}, 
``Spin torque induced dual type artificial neurons based on an in-plane magnetic tunnel junction,'' 
\textit{Appl. Phys. Lett.}, vol.~127, 202401, 2025.


\bibitem{Bradley2023}
H.~Bradley, S.~Louis, C.~Trevillian, L.~Quach, E.~Bankowski, A.~Slavin, and V.~Tyberkevych,
``Artificial neurons based on antiferromagnetic auto-oscillators as a platform for neuromorphic computing,''
\textit{AIP Advances}, vol.~13, no.~1, 2023.

\bibitem{Bradley2024}
H.~Bradley, S.~Louis, A.~Slavin, and V.~Tyberkevych,
``Pattern recognition using spiking antiferromagnetic neurons,''
\textit{Scientific Reports}, vol.~14, no.~1, p.~22373, 2024.

\bibitem{Sotnyk2025}
I.~Sotnyk and O.~Prokopenko,
``Antiferromagnetic programmable neuron: structure, training, and pattern recognition applications,''
\textit{Authorea Preprints}, 2025.

\bibitem{BradleyWithD}
H. Bradley, L. Quach, S. Louis, and V. Tyberkevych, ``Antiferromagnetic artificial neuron modeling of the withdrawal reflex,” \textit{J Comput Neurosci}, vol. 52, no. 3, pp. 197-206, July 2024.

\bibitem{ltspice}
Analog Devices, Inc., ``LTspice simulator,'' Accessed: Mar. 14, 2025. [Online].
Available: https://www.analog.com/en/resources/design-tools-and-calculators/ltspice-simulator.html



\bibitem{levakova2015review}
M.~Levakova, M.~Tamborrino, S.~Ditlevsen, and P.~Lansky, ``A review of the methods for neuronal response latency estimation,'' \textit{Biosystems}, vol.~136, pp. 23--34, 2015.

\bibitem{gerstner2002spiking}
W.~Gerstner and W.~M. Kistler, ``Spiking Neuron Models: Single Neurons, Populations, Plasticity.'' \textit{Cambridge University Press}, 2002.

\bibitem{kheradpisheh2020temporal}
S.~R. Kheradpisheh and T.~Masquelier, ``Temporal backpropagation for spiking neural networks with one spike per neuron,'' \textit{Int. J. Neural Syst.}, vol.~30, no.~6, p. 2050027, 2020.












\end{thebibliography}
\end{document}